\documentclass{aa}  

\usepackage{graphicx}
\usepackage{txfonts}
\newcommand{\abs}[1]{\ensuremath{\left|#1\right|}}
\newcommand{\e}[1]{\ensuremath{\cdot10^{#1}}}

\usepackage{amsmath}	
\newcommand{\myvector}[1]{\ensuremath{\boldsymbol{#1}}}

\usepackage{natbib}
\bibpunct{(}{)}{;}{a}{}{,} 

\begin{document}

   \title{Turbulent pressure support in galaxy clusters}

   \subtitle{Impact of the hydrodynamical solver}

   \author{Frederick Groth\inst{1}\fnmsep\thanks{\email{fgroth@usm.lmu.de}}
          \and
          Milena Valentini\inst{2,3,4,1}
          \and
          Ulrich P. Steinwandel\inst{5}
          \and
          David Vallés-Pérez\inst{6}
          \and
          Klaus Dolag\inst{1,7}
          }

    \institute{Universit{\"a}ts-Sternwarte, Fakult{\"a}t für Physik, Ludwig-Maximilians-Universit{\"a}t M{\"u}nchen, Scheinerstr. 1, 81679 M{\"u}nchen, Germany
    \and
    Astronomy Unit, Department of Physics, University of Trieste, via Tiepolo 11, I-34131 Trieste, Italy
    \and
    INAF - Osservatorio Astronomico di Trieste, via Tiepolo 11, I-34131 Trieste, Italy
    \and
    ICSC - Italian Research Center on High Performance Computing, Big Data and Quantum Computing, via Magnanelli 2, 40033, Casalecchio di Reno, Italy
    \and
    Center for Computational Astrophysics, Flatiron Institute, 162 Fifth Avenue, New York, NY 10010, USA
    \and 
    Departament d’Astronomia i Astrofísica, Universitat de València, C/Doctor Moliner 50, E-46100 Burjassot (València), Spain
    \and
    Max-Planck-Institut f{\"u}r Astrophysik, Karl-Schwarzschild-Stra{\ss}e 1, 85741 Garching, Germany
    }

   \date{Received XXX; accepted XXX}
 
  \abstract
   {The amount of turbulent pressure in galaxy clusters is still debated, especially as for the impact of the dynamical state and the hydro-method used for simulations.}
   {We study the turbulent pressure fraction in the intra cluster medium of massive galaxy clusters. We aim to understand the impact of the hydrodynamical scheme, analysis method, and dynamical state on the final properties of galaxy clusters from cosmological simulations.}
   {We perform non-radiative simulations of a set of zoom-in regions of seven galaxy clusters with Meshless Finite Mass (MFM) and Smoothed Particle Hydrodynamics (SPH). We use three different analysis methods based on: $(i)$ the deviation from hydrostatic equilibrium, $(ii)$  the solenoidal velocity component obtained by a Helmholtz-Hodge decomposition, and $(iii)$ the small-scale velocity obtained through a multi-scale filtering approach. We split the sample of simulated clusters into active and relaxed clusters.}
   {Our simulations predict an increased turbulent pressure fraction for active compared to relaxed clusters. This is especially visible for the velocity-based methods. For these, we also find increased turbulence for the MFM simulations compared to SPH, consistent with findings from more idealized simulations. The predicted non-thermal pressure fraction varies between a few percent for relaxed clusters and $\approx13\%$ for active ones within the cluster center and increases towards the outskirts. No clear trend with redshift is visible.}
   {Our analysis quantitatively assesses the importance played by the hydrodynamical scheme and the analysis method to determine the non-thermal/turbulent pressure fraction. While our setup is relatively simple (non-radiative runs), our simulations show agreement with previous, more idealized simulations, and make a step further toward the understanding of turbulence.}

   \keywords{Turbulence -- Galaxies: clusters: general -- Galaxies: clusters: intracluster medium -- Methods: numerical}

   \maketitle
%

\section{Introduction}

The Intra Cluster Medium (ICM) of galaxy clusters is a very dynamic environment and is shaped by mergers, global gas motions, and turbulence \citep{Carilli&Taylor2002, Kravtsov&Borgani2012}.
These processes act on different scales. Mergers and bulk motions directly impact the gas dynamics on large scales. Small scales, in contrast, are dominated by turbulent motions.

All of these gas motions can be described as different pressure terms contributing to the total pressure in the ICM. Besides gas motions, magnetic fields can also produce additional magnetic pressure.
In combination, the aforementioned two contributions are often summarized as non-thermal pressure opposed to the thermal pressure of the gas.

In this work, we will refer to the pressure due to small-scale turbulent motions as the turbulent pressure. The entirety of pressures except for the thermal, including turbulence, bulk motions, and possibly magnetic fields will be called non-thermal pressure.

There are a plethora of numerical and observational programs that are specifically targeted to understand the origin of the structure of the ICM. 
Turbulence is injected on large scales in merger shocks, after which it decays and cascades down to smaller scales \citep{Roettiger&Burns1999, Subramanian+2006,Mohapatra+2020a,Mohapatra+2021}.
Both simulations and observations find that ICM turbulence is subsonic, with typical velocities of a few~$100$~km/s \citep{DenHerder+2001,HitomiCollaboration+2016,HitomiCollaboration+2018}. For a sound speed of the order of~$c_s=1000$~km/s, this results in a volume filling Mach number between $\mathcal{M}=0.2$ and $0.5$, depending on the position within the clusters. 
While global measurements of turbulence in galaxy clusters can be understood within the context of the classical theory of subsonic turbulence which is supported by observations \citep[e.g. the results by][]{HitomiCollaboration+2016,HitomiCollaboration+2018}, our understanding of its origin and dissipation scales still lacks a solid base.

Various X-ray observations provide insight into ICM properties and aim to analyze ICM turbulence.
\citet{Schuecker+2004} quantify turbulence based on pressure fluctuations in the Coma cluster. They find that turbulence is well described by a Kolmogorov power spectrum, with an upper limit of the turbulent pressure of $10\%$ of the total pressure.
Similarly, \citet{Zhuravleva+2019} study density fluctuations from X-ray observations and quantify viscosity in the ICM. They derive velocities up to around a few $100$~km/s, closely following the expected Kolmogorov scaling for subsonic turbulence.
Using spectroscopic lines with XMM-Newton, \citet{Gatuzz+2023a} confirm the Kolmogorov-like slope with driving scales of $10-20$\,kpc in Virgo.

Exploiting spectroscopically resolved lines to derive velocities, \citet{HitomiCollaboration+2016,HitomiCollaboration+2018} perform detailed measurements in the Perseus cluster, yielding velocities between $100-200$~km/s and a turbulent pressure support of only $4\%$ compared to the total pressure. This value is at the lower end compared to many other results cited in this work. As Perseus also shows a cool core, it is often classified as relaxed in the center where the turbulence is measured, despite the cluster showing sloshing motions and an active galactic nucleus (AGN). As this is a single system, it is unclear from the aforementioned work if there is any dependence on the dynamical state.
The result is still consistent with the upper limits of the spectroscopic measurements by XMM-Newton \citep{DenHerder+2001} of $200-600$~km/s.

An alternative method is used by \citet{Eckert+2019}. For the X-COP sample, they quantify the non-thermal pressure based on the deviation from hydrostatic equilibrium (HE) and find values between $2\%$ and $15\%$, depending on radius and cluster.

Different observations indicate that the amount of turbulence should depend on the dynamical state of the system. One indirect tracer of turbulence is the existence of a radio halo, which requires turbulent re-acceleration. \citet{Cassano+2010,Cuciti+2015,Cuciti+2021b} found that merging systems typically host such a radio halo, while relaxed systems do not.

Alternative insight can be gained from simulations of cosmological boxes or zoom-in regions with adequate resolution. One main advantage is the access to the velocity data for every resolution element in the simulation. Nevertheless, it remains difficult to extract turbulence directly, as this requires a good estimate of the bulk flow.

Based on the velocity dispersion, \citet{Lau+2009} find a turbulent pressure support between $6-15\%$, increasing with radius, in relaxed systems and higher values between $9-24\%$ in unrelaxed systems.
Using a slightly different approach, i.e. separating the smooth gas component from clumps, and computing median instead of mean properties, \citet{Zhuravleva+2013c} find consistent results. They find increased root-mean-square (rms) velocities in active clusters of $\approx0.7 \, c_s$ in units of the sound speed, compared to $\approx0.4 \, c_s$ in relaxed clusters.

In a series of papers, \citet{Vazza+2009, Vazza+2012a, Vazza+2017, Vazza+2018} explore turbulence in galaxy clusters simulated with the \textsc{ENZO} code \citep{Bryan+2014} featuring adaptive mesh refinement \citep[AMR;][]{Berger&Colella1989}. They introduce the multi-scale filtering technique, which decomposes the total velocity into bulk and turbulent motions, predicting a turbulent pressure support around $10\%$. In addition, they show a strong dependence on whether a constant or variable filtering length is used. Such a dependence is expected because the filtering length allows us to focus on motions only on the length scales defined by the size of turbulent eddies, ignoring any bulk motions on larger scales that can increase the kinetic energy budget.

\citet{Biffi+2016} use a modern SPH implementation with a high-resolution shock-capturing method and a high-resolution entropy-increasing diffusion scheme (also known as artificial viscosity and conduction), with an additional stabilization of the method against the tensile instability using a high-order kernel for their simulations. They find overall high devitations from HE around $10-20\%$, with a higher deviation for more disturbed systems.
This is slightly higher than other values from previously quoted references in the introduction, but still consistent.
In addition, not all of this deviation from HE will be attributed to turbulence, instead this is only an upper limit. 

A direct comparison between simulations and observations has been made by \citet{Sayers+2021}. Using the ``Clump3d'' method, they combine several observations to de-project the cluster and derive a non-thermal pressure based on the deviation from HE.
Even when analyzed with the same method, simulations and observations show a significant disagreement in the non-thermal pressure fraction. It reaches $\approx13\%$ for simulations, while it is consistent with zero for observations.
In addition, \citeauthor{Sayers+2021} find no dependence on the dynamical state.

Overall, the interpretation of the non-thermal pressure depends on the analysis method. Many strategies based on the deviation from HE include the effect of bulk motions, magnetic fields, cooling flows, turbulence, and everything not attributed to the thermal origin.
Also rotational patterns can affect the non-thermal pressure support \citep{Biffi+2011}.
This makes the comparison between different results depending on subtleties and motivates the need for more robust methods.
More direct methods such as the multiscale-filtering technique return instead the actual turbulent pressure, filtering out motions on larger scales.

Derived properties of ICM turbulence depend not only on the analysis method but also on the numerical setup. 
Detailed comparisons on different hydro-methods to simulate subsonic turbulence have been made mainly in idealized boxes \citep[compare, e.g.,][]{Kitsionas+2009,Price&Federrath2010,Padoan+2007,Bauer&Springel2012,Price2012}. These works show that especially the power spectrum can be significantly impacted by the choice of the hydro scheme.

In cosmological simulations, several works studied the impact of individual numerical parameters. A too high artificial viscosity can significantly reduce the amount of turbulence \citep{Dolag+2005b}. Also, artificial conductivity reduces turbulence, as a result of increased gas stripping \citep{Biffi&Valdarnini2015a}.

In this work, we want to extend the comparison of different hydrodynamical methods from idealized simulations to cosmological environments. In particular, we want to study the differences between Meshless Finite Mass \citep[MFM;][]{Lanson&Vila2008,Lanson&Vila2008a} and Smoothed Particle Hydrodynamics \citep[SPH;][]{Springel&Hernquist2002} as two different hydro methods on the resulting turbulence in the ICM.
To this end, we use non-radiative, hydrodynamical zoom-in simulations of galaxy clusters to have a clean setup that allows us to produce robust results to compare MFM and SPH.
In addition, we want to quantify the impact of the analysis method, as well as of the dynamical state of the cluster.

The paper is organized as follows.
In Sec.~\ref{sec:simulation_setup} we describe the code and simulation setup, followed by a description of the different methods used to analyze the simulations in Sec.~\ref{sec:analysis_methods}.
The general dynamical and thermodynamic properties of the clusters used in this work are described in Sec.~\ref{sec:cluster_properties}. To gain insight into the turbulence, we start with a general analysis of the velocity structure in Sec.~\ref{sec:velocity_structure}, and present the resulting turbulent pressure support in Sec.~\ref{sec:turbulent_pressure}.
Our findings are discussed in Sec.~\ref{sec:conclusions}

\section{The simulations} \label{sec:simulation_setup}
\subsection{OpenGadget3}
The simulations are performed with the hydrodynamical cosmological simulation code \textsc{OpenGadget3}. It is originally based on \textsc{Gadget-2} \citep{Springel+2001,Springel2005}. 
Gravity is calculated with an Oct-Tree-Particle Mesh (PM) approach \citep{Xu1995,Springel2005,Springel+2021}. Hydrodynamical forces are calculated either using modern SPH \citep{Springel&Hernquist2002} including artificial viscosity as formulated by \citet{Beck+2016} and artificial conductivity as formulated by \citet{Price2008}. Alternatively, the MFM hydro-solver is used, with the implementation presented by \citet{Groth+2023c}.

$295$ neighbors are used for the SPH calculations with a Wendland C6 kernel \citep{Wendland1995,Dehnen&Aly2012}, while only $32$ neighbors with a cubic spline kernel \citep{Monaghan&Lattanzio1985} are best suited for MFM. This leads to an effectively higher spatial resolution for the hydrodynamical solver defined by the smoothing length $h$ using MFM compared to SPH by a factor of $\approx 3$.

The \textsc{Subfind} substructure finder \citep{Springel+2001,Dolag+2009} and a shockfinder \citep{Beck+2016a} are run on-the-fly.

\subsection{Dianoga suite}
We simulate seven massive galaxy clusters from the Dianoga suite of zoom-in regions \citep{Bonafede+2011a}. The background cosmological evolution follows a flat $\Lambda$CDM cosmology with $\Omega_{\rm m}=0.24$, $\Omega_{\rm b}=0.04$, $h=0.72$ and $\sigma_8=0.8$.
The mass resolution is $M_{\rm dm}=10^{9}h^{-1}$~M$_{\sun}$, $M_{\rm gas}=1.6\e{8}h^{-1}$~M$_{\sun}$.
The gravitational softening corresponds to $3.75$~$h^{-1}$kpc for gas and $11.25$~$h^{-1}$kpc for DM particles.

All selected clusters have a mass larger than $10^{15}h^{-1}$~M$_{\sun}$.
Their names and masses are listed in Table~\ref{tab:dianoga_clusters}.
\begin{table}[tbp]
    \caption{\label{tab:dianoga_clusters} List of some general properties of the galaxy clusters analyzed in this study.}
    \centering
    \begin{tabular}{c|cc|ccc|c}
        Cluster ID & $M_{200}$ & $M_{200}^\text{gas}$ & $R_{200}$ & $R_{500}$ & $R_{2500}$ & active/  \\
        & \multicolumn{2}{|c|}{[$10^{15}h^{-1}{\rm M}_{\sun}$]} & \multicolumn{3}{|c|}{[$h^{-1}$Mpc]} & relaxed \\ 
        \hline
        {\bf g12}12639 & $1.9$ & $0.28$ & $3.2$ & $2.2$ & $0.5$ & $~61\%/~39\%$ \\
        {\bf g14}83463 & $1.6$ & $0.24$ & $3.1$ & $2.2$ & $0.4$ & $100\%/~~0\%$ \\
        {\bf g165}7050 & $1.9$ & $0.29$ & $3.2$ & $2.1$ & $0.5$ & $~47\%/~53\%$ \\
        {\bf g168}0241 & $1.7$ & $0.26$ & $3.1$ & $2.0$ & $0.4$ & $~89\%/~11\%$ \\
        {\bf g19}87669 & $1.9$ & $0.29$ & $3.2$ & $2.3$ & $0.5$ & $~96\%/~~4\%$ \\
        {\bf g55}03149 & $1.6$ & $0.25$ & $3.1$ & $2.1$ & $0.5$ & $~88\%/~12\%$ \\
        {\bf g63}48555 & $1.7$ & $0.28$ & $3.1$ & $2.1$ & $0.5$ & $~~0\%/100\%$
    \end{tabular}
    \vspace{4 pt}
    \tablefoot{The bold part of the identification number is used as an abbreviation in the text. The last column gives the fraction of the evolution time since $z=0.43$ that the cluster is relaxed/active. The dynamical state was determined by the method described in Sec.~\ref{sec:dynamical_state}. All properties agree between MFM and SPH in the given accuracy and differences are $\lesssim 1\%$.}
\end{table}
The selection includes active clusters with high merger activity and relaxed clusters that undergo mainly smooth accretion or minor mergers. This diversity allows us to study the effect of the dynamical state and compare results to other simulations and observations of populations of different clusters.

Every cluster was simulated twice, once with the MFM solver, and once with SPH. The main properties such as $R_{200}$ and $M_{200}$ quoted in Tab.~\ref{tab:dianoga_clusters} are consistent between the methods. Also, the classification between active and relaxed systems including the fraction of their evolution they stay active or relaxed does not change when different hydro-solvers are used. Overall, differences between the methods are $\lesssim1\%$ for all quantities and thus the values cited in Tab.~\ref{tab:dianoga_clusters} are identical within the given accuracy.

For each simulation, we create 46 snapshots from redshift $z=0.43$ until redshift zero, which we use for time averaging and studying their evolution.

\begin{figure}
    \centering
    \includegraphics[width=\columnwidth]{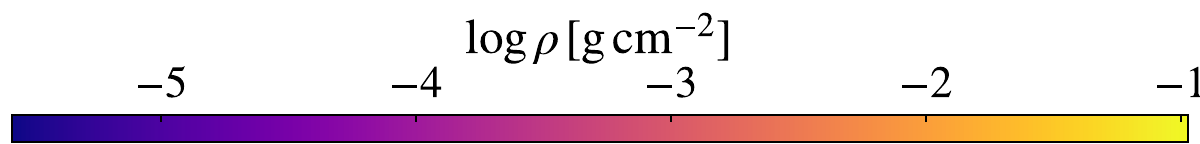}
    \includegraphics[width=\columnwidth]{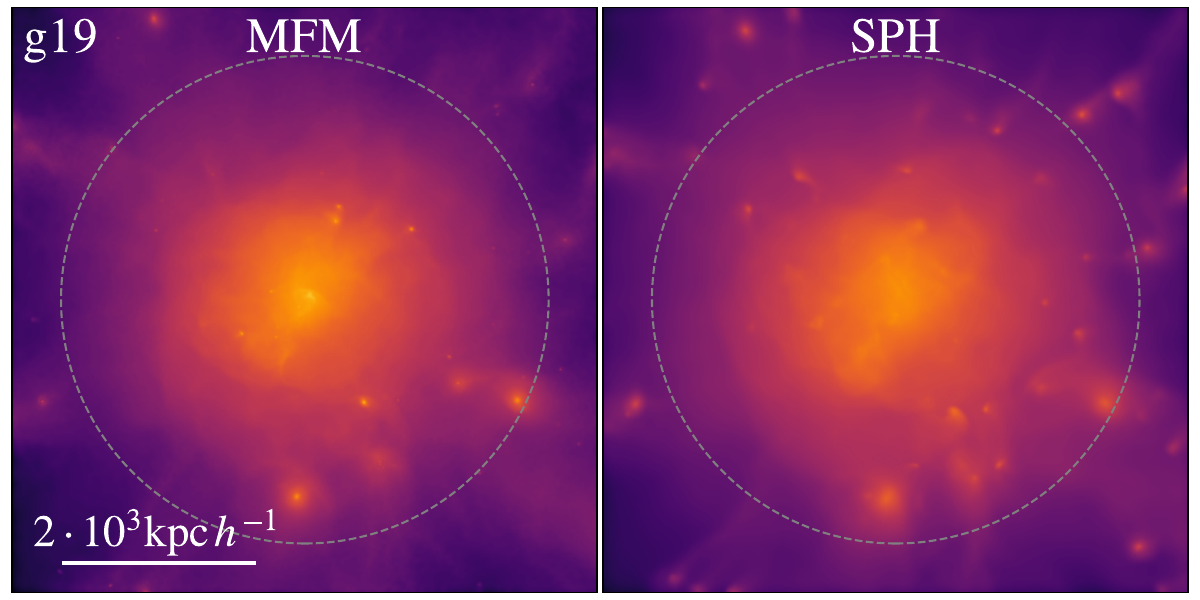}
    \includegraphics[width=\columnwidth]{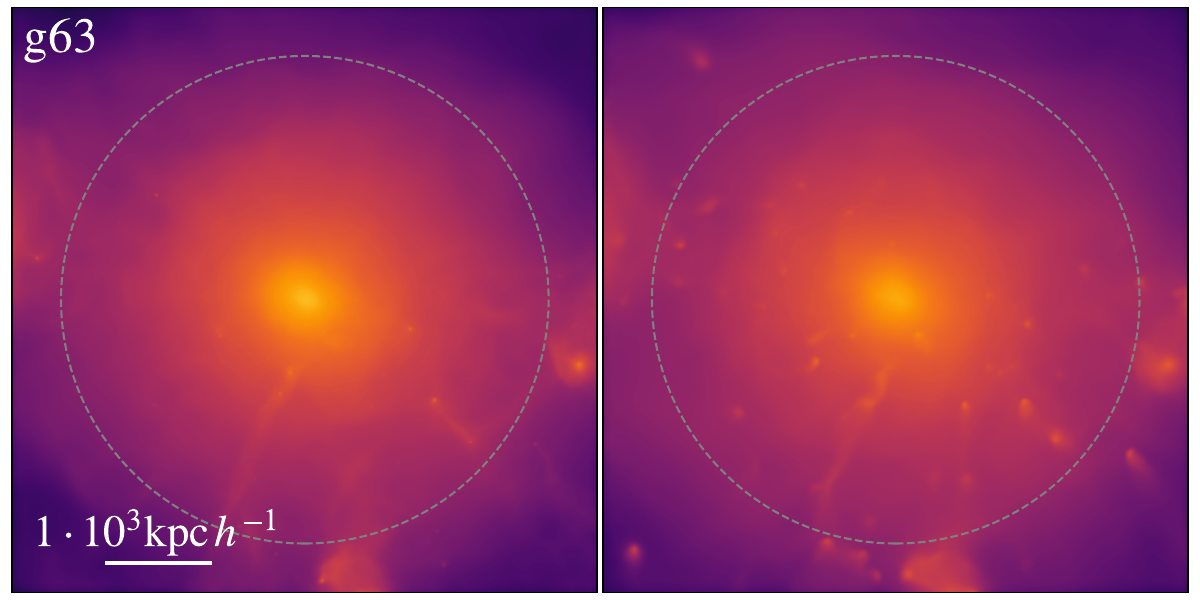}
    \caption{\label{fig:maps} Projected gas density maps for g19 and g63 as examples of one more active and one more relaxed cluster analyzed in this work at redshift $z=0$. The dashed circle denotes $R_{\rm vir}$. The upper maps have a size of $\Delta x=\Delta y\approx3064$~kpc\,$h^{-1}$, the lower maps $\Delta x=\Delta y\approx2797$~kpc\,$h^{-1}$.}
\end{figure}

\section{Analysis methods} \label{sec:analysis_methods}

\subsection{Dynamical states} \label{sec:dynamical_state}
To classify the dynamical state of a galaxy cluster, we follow the method described by \citet{Cui+2017,Cui+2018}. We use two complementary criteria, where clusters are classified as active if at least one of them is satisfied.
The first criterion is based on the shift of the center of mass (com) with respect to the cluster center defined by the minimum potential, where active clusters have an offset of at least
\begin{align}
    \abs{\myvector{r}_{\text{min pot}} - \myvector{r}_{\text{com}}} \geq~& 0.04 \, R_{200}.
\end{align}
In addition, clusters for which the mass in substructures exceeds
\begin{align}
    M_{\text{sub}} \geq~& 0.1 M_{200}
\end{align}
are considered active.
The cluster is classified as relaxed if none of the criteria are satisfied.
The minimum potential, as well as the mass enclosed in substructures, is found using \textsc{Subfind} \citep{Springel+2001,Dolag+2009}. The center of mass is calculated for all gas and dark matter (DM) particles within $R_{200}$\footnote{All quantities ($R_{200}$, $M_{200}$, etc) are defined with respect to to the mean mass density of the universe} from the \textsc{Subfind} center.

Both criteria are directly related to major mergers. The first one aims to capture the offset and sloshing of the gas within the DM potential shortly after the merger. Also massive substructures can offset the global mass distribution. The second criterion detects the infalling halo of an ongoing merger more directly.

Alternative criteria to classify the dynamical state would be possible, e.g. based on the virialization \citep[compare, e.g.,][]{Cui+2018}. We found, however, that these closely follow the two criteria used in this work, such that they are sufficient to classify the dynamical state of the system.

\subsection{Clump3d analysis}
The first option to calculate the non-thermal pressure contribution closely follows the method described by \citet{Sayers+2021}.
In the first step, the ellipticity of the gas and total matter distribution is calculated according to the framework described by \citet{Fischer+2022,Fischer&Valenzuela2023} to account for deviations from spherical symmetry. The derived axes are used to define an elliptical radial coordinate used in all the fits.

The global density profile of gas plus DM is fit with a Navarro-Frenk-White (NFW) profile \citep{Navarro+1997}.
The gas density is fit with a modified beta-model, and its temperature with a modified broken power law \citep{Vikhlinin+2006}.
The total pressure is obtained from these fits using the HE equation: 
\begin{align}
    \nabla P_{\rm tot} =~& -\rho\nabla\Phi_{\rm mat}.
\end{align}
The thermal pressure is calculated directly from the gas density $\rho_{\rm gas}$ and internal energy $u_{\rm gas}$ of all particles as:
\begin{align}
    P_{\rm therm} =~& \left(\gamma-1\right)\rho_{\rm gas} u_{\rm gas} \label{eq:Pth}
\end{align}
with adiabatic index $\gamma=5/3$.
We use 40 radial bins equally distributed in log space between $0.01 \, R_{200}$ and $1.1 \, R_{200}$ to compute these profiles, using a logarithmic mean over the particles within each bin.
The non-thermal pressure is given by the deviation from the assumption of HE. It is computed as the difference between the total and thermal pressure $P_{\rm nt} = P_{\rm tot}-P_{\rm therm}$, limited to values greater than zero.
Finally, the time and cluster-to-cluster average is computed with the radial coordinate normalized to $R_{200}$.

The non-thermal pressure derived from this method includes contributions from turbulence and bulk motions. Depending on which physics is included, it could also contain e.g. magnetic pressure. In addition, strong shocks after recent mergers could impact the derived non-thermal pressure.

\begin{figure*}
    \centering
    \includegraphics[width=\textwidth]{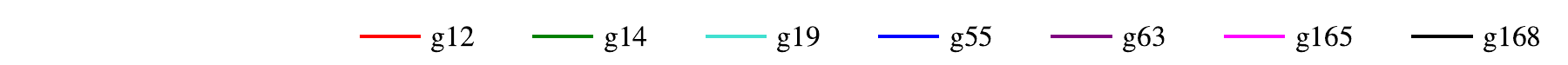}
    \includegraphics[width=\textwidth]{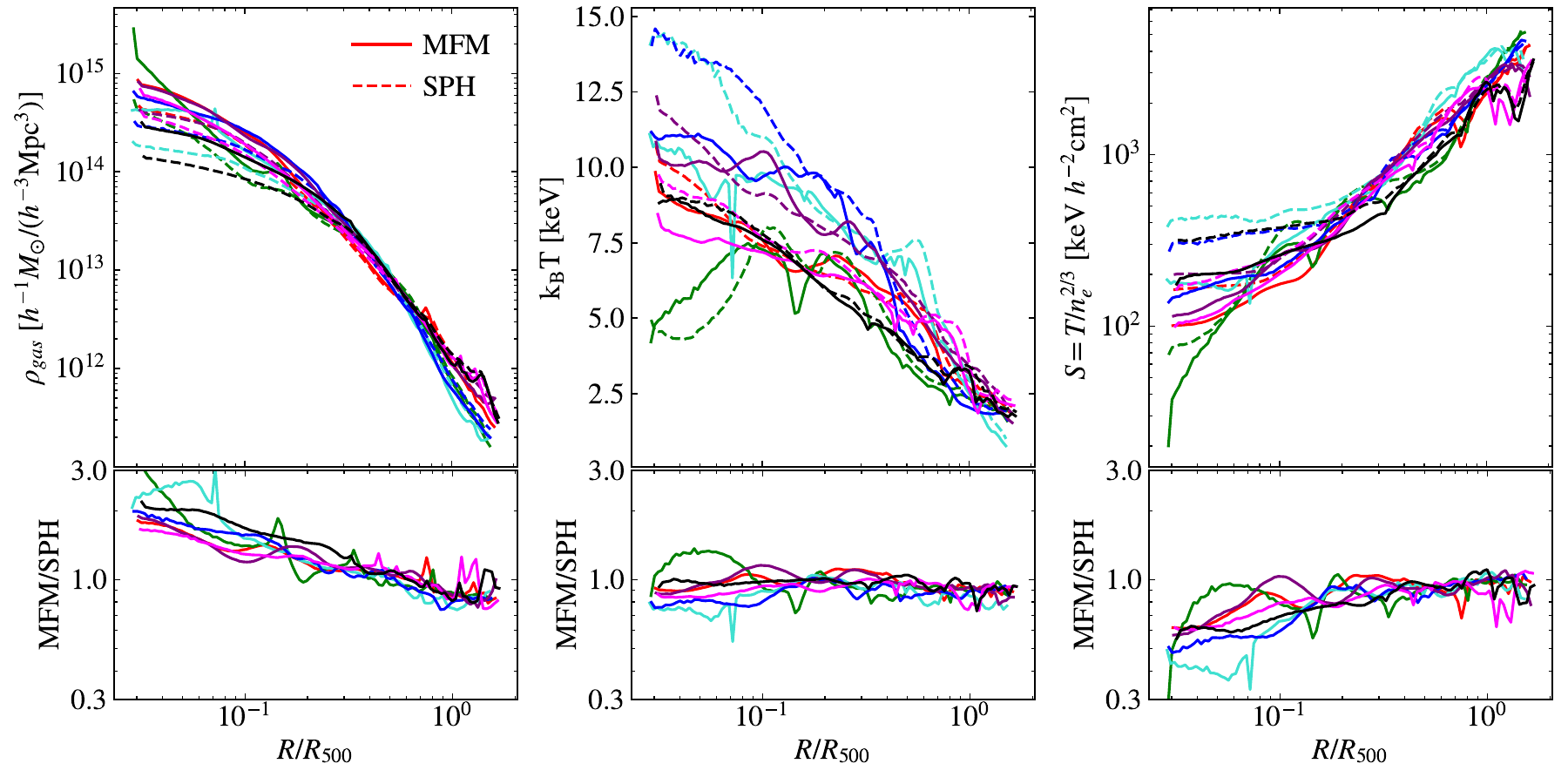}
    \caption{\label{fig:radial_profiles} Radial gas density, temperature, and entropy profiles for the clusters analyzed in this work at redshift $z=0$. Lower panels show the ratio between MFM and SPH simulations for each cluster.}
\end{figure*}

\subsection{Vortex analysis -- Helmholtz decomposition}
The second method to calculate the non-thermal pressure fraction uses the velocity data from the simulation more directly.

In particular, the solenoidal component of the velocity is associated with turbulence, while compressive motions are associated with shocks and bulk motion, following the results of \citet{Vazza+2017} who find that the solenoidal component of turbulence dominates.
In principle, the solenoidal component also contains components of motions at larger scales. Nevertheless, it tends to be more isotropic, thus behaving more like an additional pressure contribution compared to the compressive part (mainly radial; cf. figure~7 in \citealp{Valles-Perez+2021b}).
Also compressive velocity components can be associated with turbulence, and the mixture between solenoidal and compressive can vary depending on the type and driving of turbulence \citep{Federrath+2010,Federrath+2021}. These results have been found for supersonic turbulence. For subsonic turbulence within galaxy clusters a predominance of the solenoidal component has been found \citep{Ryu+2008,Vazza+2014,Vazza+2017,Wang&He2024a}. Overall, the solenoidal component is only an indirect tracer of turbulence, and we will compare the performance in more detail in the remainder of the paper.

We use the \textsc{vortex-p} code developed by \citet{Valles-Perez+2024a} to perform a Helmholtz-decomposition \citep{Valles-Perez+2021b}. The total gas velocity is split into compressive and solenoidal components which are derived via a scalar potential $\phi$ and vector potential $\myvector{A}$, respectively, leading to a unique decomposition
\begin{align}
    \myvector{v} =~ \myvector{v}_{\rm compressive} + \myvector{v}_{\rm solenoidal},\\
    \myvector{v}_{\rm compressive} =~ -\nabla\phi,~~~~\myvector{v}_{\rm solenoidal} =~ \nabla\times\myvector{A}.
\end{align}
The potentials are found as solutions to the elliptic partial differential equations
\begin{align}
    \nabla^2\phi=~ -\nabla\cdot\myvector{v},~~~~\nabla^2\myvector{A}=~-\nabla\times\myvector{v}.
\end{align}

Internally, \textsc{vortex-p} assigned the SPH data onto an ad-hoc set of nested AMR grids, with higher resolution in regions of higher particle number density, especially in the cluster center and within substructures.

We run the decomposition in a region of $50h^{-1}\rm{Mpc}$ sufficiently large to contain the virialized region of the main cluster and not be influenced by boundary effects.
The base grid has a resolution of $N_x=128$ and a maximum of $n_l=6$ refinement levels. Cells at any refinement level containing more than $n^{\rm refine}_{\rm part}=8$ particles are set to refine, thus producing a quasi-Lagrangian refinement. These settings result in a peak resolution of $\Delta x_6 \approx 6 \rm{kpc}$ which is on the same order as the minimum smoothing length. A cubic spline kernel is used to interpolate the MFM simulations, while a Wendland C6 kernel is used for the SPH runs. Interpolations in the decomposition use the same neighbor number as in the cosmological simulation they are applied to. Further details on the setup of the cosmological simulations are provided in Sec.~\ref{sec:simulation_setup}.
The solenoidal velocity is then mapped back from the internal AMR grid to the original particle positions. This can introduce some smoothing and small errors, which are, however, of the same order as the errors involved in the initial grid assignment (see \citealp{Valles-Perez+2024a}, their figures 2 and 3). 

The turbulent pressure within $40$ elliptical shells is calculated directly from the particle-based velocity data:
\begin{align}
    P_{\text{turb}} =~& \frac{1}{3}\rho v_{\text{sol}}^2. \label{eq:Pturb}
\end{align}
Also the thermal pressure is obtained directly from the gas properties according to Eq.~(\ref{eq:Pth}).
The total pressure is the sum of both. All properties are calculated as mass-weighted mean.

\subsection{Vortex analysis -- multiscale filtering}
Finally, \textsc{vortex-p} can also perform a Reynolds decomposition to split the bulk component of the velocity field from the turbulent contribution. Following the initial idea by \citet{Vazza+2012a,Vazza+2017}, we perform a multiscale filtering approach. The details of the implementation and the extension of the method to AMR have been described by \citet{Valles-Perez+2021a}. 

The outer scale of turbulence is constrained iteratively for each cell center.
A lower bound $L_0=3\Delta x_l$ depends on the local resolution of the ad-hoc AMR grid (i.e., the refinement level $l$). Then, the filtering scale is iteratively increased until the turbulent velocity converges, indicating that longer spatial scales around the given point no longer contain more kinetic energy, and hence the outer scale of the inertial range has been reached.
To avoid divergent behavior at discontinuities, the iteration also stops if a shocked cell with Mach number $\mathcal{M}\geq2.0$ enters the integration domain. The resulting filtered mean velocity corresponds to the bulk motion, such that the turbulent velocity remains as the difference between total and filtered mean velocity.
As for the solenoidal velocity, the filtered velocities are mapped back to the particle position.

This decomposition offers the most direct way of measuring the turbulent velocity.
The turbulent pressure is calculated according to Eq.~(\ref{eq:Pturb}), replacing the solenoidal velocity with the filtered one.

\section{Cluster properties} \label{sec:cluster_properties}
To better understand the set of clusters, we first analyze some of their main properties.
The projected gas density maps of two clusters representative of both more active and more relaxed clusters are shown in Fig.~\ref{fig:maps}. Maps of the remaining clusters are shown in the Appendix. Overall, the galaxy clusters simulated with MFM have general properties which are similar to those of their analogous SPH version, by visual inspection. The same is true for the location of substructures, as the large-scale evolution is dominated by gravity. Also the accretion history of the gas is almost identical, with only minor timing differences.

Differences are visible on smaller scales.
Small structures tend to be destroyed earlier for MFM, leading to less substructures in the cluster. We do not find a statistically significant difference in the subhalo gas mass function between MFM and SPH within $R_{200}$ of the cluster, as it is highly dominated by uncertainties in the substructure finder for such small halos. For MFM, we even find an overall higher number of subhalos. Nevertheless, \textsc{Subfind} struggles to find the gas content and the differences are minor when manually calculating the mass of the gas enclosed within the substructure. However, the smaller number of substructures in MFM simulations appears to be a consistent trend by visual inspection of all surface density plots.
This observation is a sign that MFM mixes gas more efficiently and is more dissipative, consistent with more idealized simulations shown by \citet{Groth+2023c}.
In addition, from visual inspection, the diffuse volume-filling ICM appears to have more turbulent density fluctuations, visible mainly in the active cluster g19.

Cluster g19, classified as an active cluster at almost all redshifts, has more and larger substructures and ongoing mergers. In contrast, more relaxed clusters, such as g63, have less substructures and undergo mainly smooth accretion. Their overall shape looks much smoother and rounder.

The first panel shows the gas density $\rho_{\rm gas}$, the second panel the gas temperature in terms of $k_{\rm B}T$, and the third panel the entropy $S=T/n_e^{2/3}$ profile. The electron density $n_e$ is calculated assuming full ionization.
In the outskirts, the radial profiles shown in Fig.~\ref{fig:radial_profiles} are fully consistent between MFM and SPH.
As was already visible in the maps, differences are present in the center.
In particular, MFM tends to produce slightly cooler and denser cores compared to SPH.
The central entropy is lower for MFM compared to SPH by a factor of $\approx2$, leading to the density higher by a factor $\approx2$. This difference can be explained by MFM having more mixing and slightly higher numerical diffusivity, as discussed by \citet{Groth+2023c} in the more idealized case of the hydrostatic sphere.

Most of the clusters have a hot core, visible both in temperature and in entropy. Only the g14 cluster has a cool core with a decrease in temperature in the center. This is most likely a transition state of the ongoing strong merger activity. In the surface density plot in App.~\ref{app:surface_density_plots}, even two distinct density peaks in the center are visible as a result of a recent major merger. This can significantly impact all radial profiles for this particular cluster.

\begin{figure}
    \centering
    \includegraphics[width=\columnwidth]{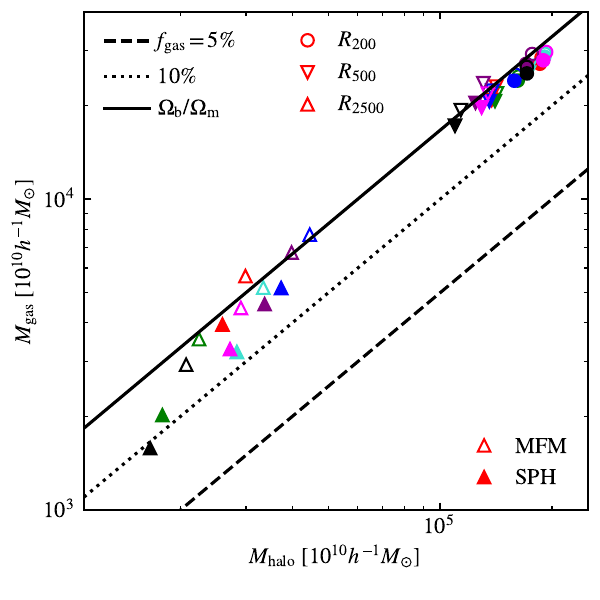}
    \caption{\label{fig:gas_halo_mass} Gas mass to halo mass relation for the clusters evaluated inside different radii at redshift $z=0$. Lines indicate constant gas fractions $f_{\rm gas}$. Same colors for each cluster as in Fig.~\ref{fig:radial_profiles}.}
\end{figure}

The gas mass to halo mass relation for the different clusters evaluated inside different radii $R_{200}$, $R_{500}$, and $R_{2500}$ is shown in Fig.~\ref{fig:gas_halo_mass}.
MFM and SPH lead to almost identical results when evaluating the masses inside larger radii (i.e., $R_{200}$ and $R_{500}$), as the global structure does not change. The gas content reaches roughly the cosmological one $\Omega_{\rm b}/\Omega_{\rm m}\approx 16.7\%$. Strong clustering in the cluster center can lead to gas contents even above the cosmological value observed for a few clusters.

Slight differences appear in the innermost region inside $R_{2500}$, where MFM generally leads to larger masses, consistent with the increase in density found in the radial profiles.
The gas content ranges from $10\%$ to approximately the cosmological value of $16.7\%$. 
While this is slightly larger than typical observation results around $5-13\%$ \citep{Vikhlinin+2006,David+2012}, differences can be explained by the absence of AGN feedback processes, which would reduce the gas fraction, especially at smaller radii \citep{McNamara+2000,Churazov+2000,Eckert+2021}, and also star formation which reduces the gas as it is transformed into stars.

\begin{figure}
    \centering
    \includegraphics[width=\columnwidth]{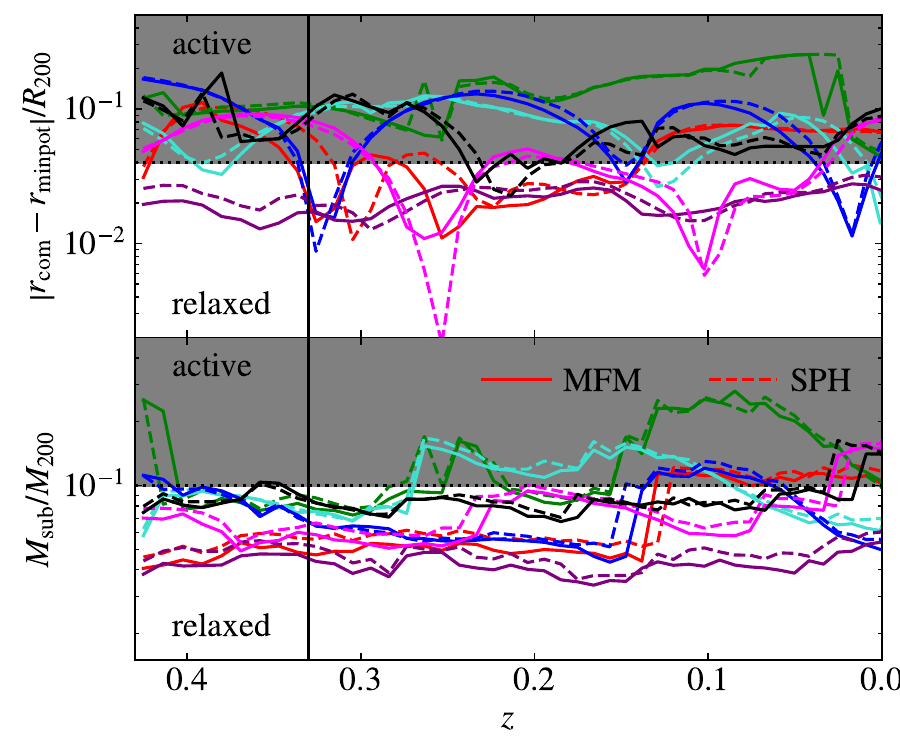}
    \caption{Evolution of the center offset and substructure-mass over time of all clusters analyzed in this work. As described in Sec.~\ref{sec:dynamical_state}, a cluster is classified as relaxed only if both lines are below the two dotted thresholds, and otherwise as active. Same colors for each cluster as in Fig.~\ref{fig:radial_profiles}.}
    \label{fig:dynamical_states}
\end{figure}

The clusters show a variety of dynamical states as shown in Fig.~\ref{fig:dynamical_states}. As described in Sec.~\ref{sec:dynamical_state}, clusters are classified as relaxed if both, offset of the center of mass and mass enclosed in substructures, are below the threshold marked by the gray-shaded region, while they are classified as active otherwise.
Sloshing can make the center of mass approach the center defined by the minimum potential at some redshifts, but still have a non-zero relative velocity. Our second criterion of the mass enclosed in substructures ensures that the clusters can also be classified as active under such a condition.

Some clusters such as g14 (green) or g63 (purple) remain active or relaxed, respectively, at all redshifts.
Other clusters such as g55 (dark blue) can change their dynamical state over time, becoming active after larger mergers, but becoming relaxed again soon after.

There are only minor differences between the two hydro-methods. In particular, the mass enclosed in substructures is slightly smaller for MFM compared to that of SPH, but this does not change the dynamical state classification.
Overall, using two criteria allows for a stable classification of the dynamical state.

\section{Velocity structure} \label{sec:velocity_structure}

\begin{figure*}
    \centering
    \includegraphics[width=\linewidth]{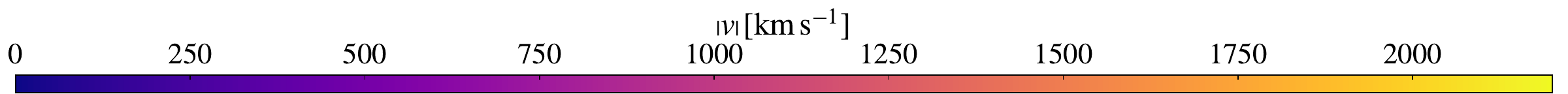}\\
    \includegraphics[width=\linewidth]{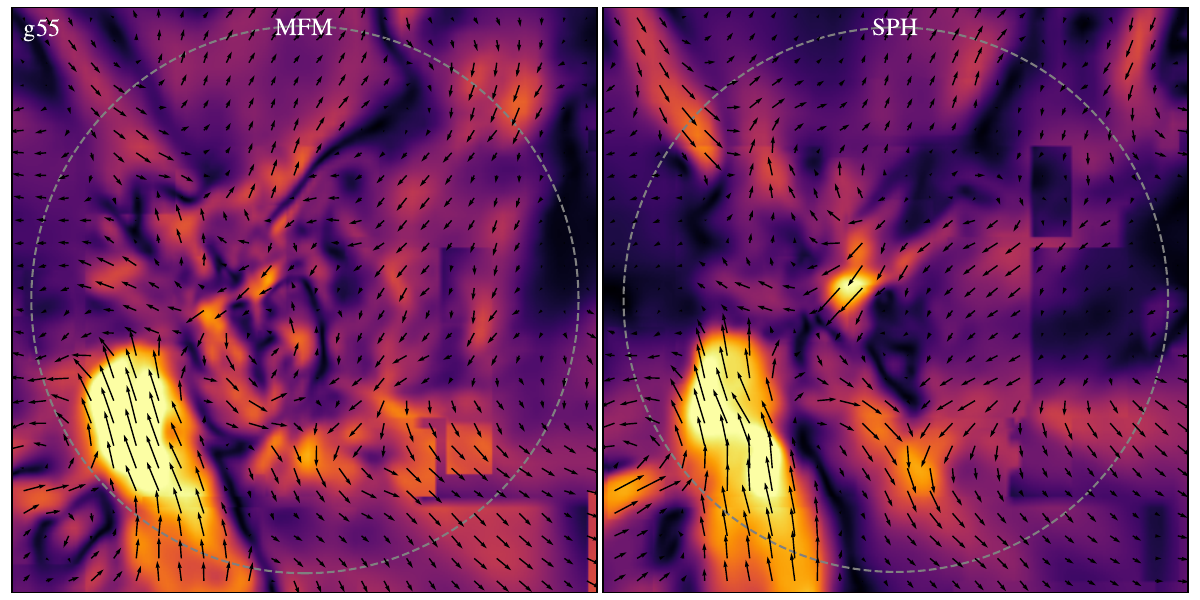}\\
    \includegraphics[width=\linewidth]{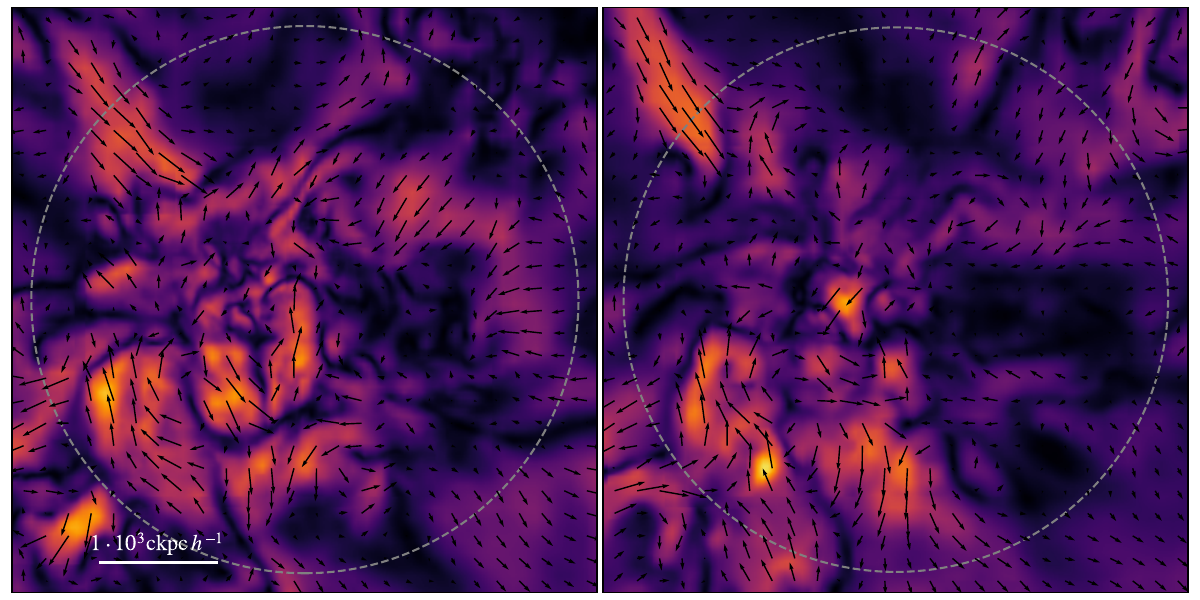}
    \caption{Slice through the g55 cluster at redshift $z=0$, showing the rotational component of the velocity in the upper panel and the multi-scale filtered velocity in the lower panel, each comparing MFM and SPH. The color indicates the absolute value of the solenoidal/filtered velocity, while the quivers show the direction. The dashed circle marks $R_{\rm vir}$.}
    \label{fig:vel_rot_filtered}
\end{figure*}

A more turbulent structure of the ICM for MFM compared to SPH is already visible from the surface density fluctuations.
A more direct tracer is the velocity structure. In the upper part of Fig.~\ref{fig:vel_rot_filtered} we show the solenoidal velocity component in a slice through the g55 cluster.
As expected, the global structure is qualitatively very similar between MFM and SPH. A strong increase can be found within a small region at the lower left. The total velocity shows that this region also has a very high infall velocity in general.

Quantitatively MFM leads to higher velocities on average, and regions of large velocities are more extended. In addition, MFM has finer structures in the central region, which can be a result of more turbulent structure and also the effectively higher resolution due to the smaller neighbor number.

\begin{figure*}
    \centering
    \includegraphics[width=\textwidth]{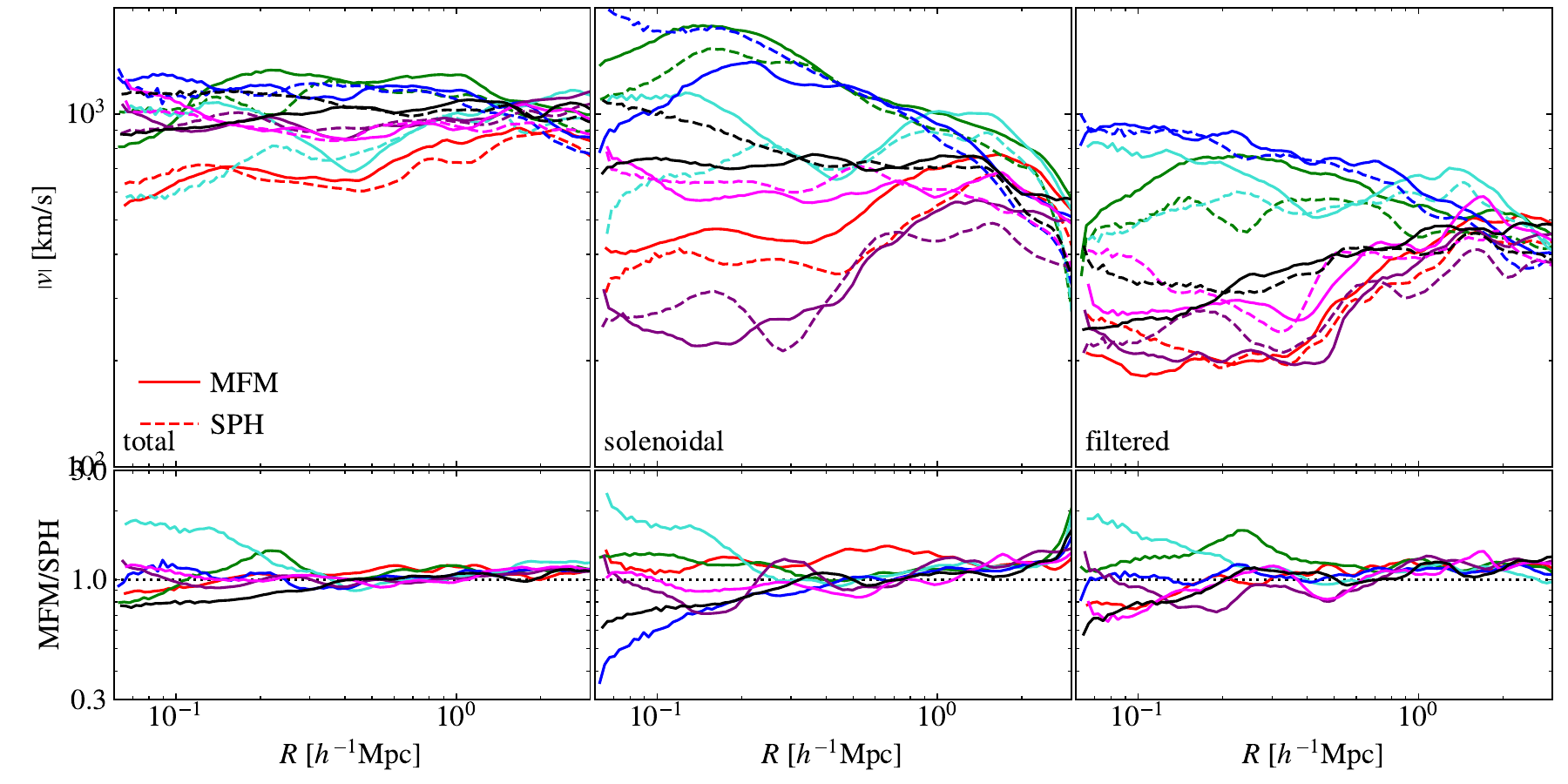}
    \caption{Velocity profiles of all the simulated galaxy clusters, at redshift $z=0$. Same colors for each cluster as in Fig.~\ref{fig:radial_profiles}. The lower panels show the ratio between MFM and SPH simulations for each cluster.}
    \label{fig:velocity}
\end{figure*}

The even more direct tracer for turbulence is the filtered velocity, shown in the lower panels of Fig.~\ref{fig:vel_rot_filtered}.
The patterns are similar to the solenoidal component, especially in the center. Nevertheless, the filtered velocity has a much lower maximum and does not show a strong increase within the accreting region.
These differences persist and become clearer when averaging the absolute value within spherical shells, as shown in Fig.~\ref{fig:velocity}.
Differences between MFM and SPH are small, and substantial scatter between different clusters can be observed in the center. A mild increase of the order of $10\%$ for MFM compared to SPH is visible in the outskirts for all panels, most pronounced for the filtered and solenoidal components. In the center, the differences are less obvious due to the strong scatter between different clusters, and within the scatter, it is consistent with all panels being identical between the hydrodynamical methods. More detailed conclusions can be drawn in Sec.~\ref{sec:turbulent_pressure}, when also averaging over different redshifts studying the resulting pressure.

While turbulence is mainly solenoidal \citep[compare, e.g.,][]{Vazza+2017}, also other solenoidal motions can be present in the ICM. Thus, the solenoidal component is a possible tracer of turbulence, but it tends to overestimate the turbulent velocity compared to the filtered one for almost every cluster. It is only an indirect tracer, and should be used with caution. Nevertheless, as we will see later, it still works on average reasonably well.

\section{Turbulent pressure fractions} \label{sec:turbulent_pressure}

\begin{figure*}
    \centering
    \includegraphics[width=\linewidth]{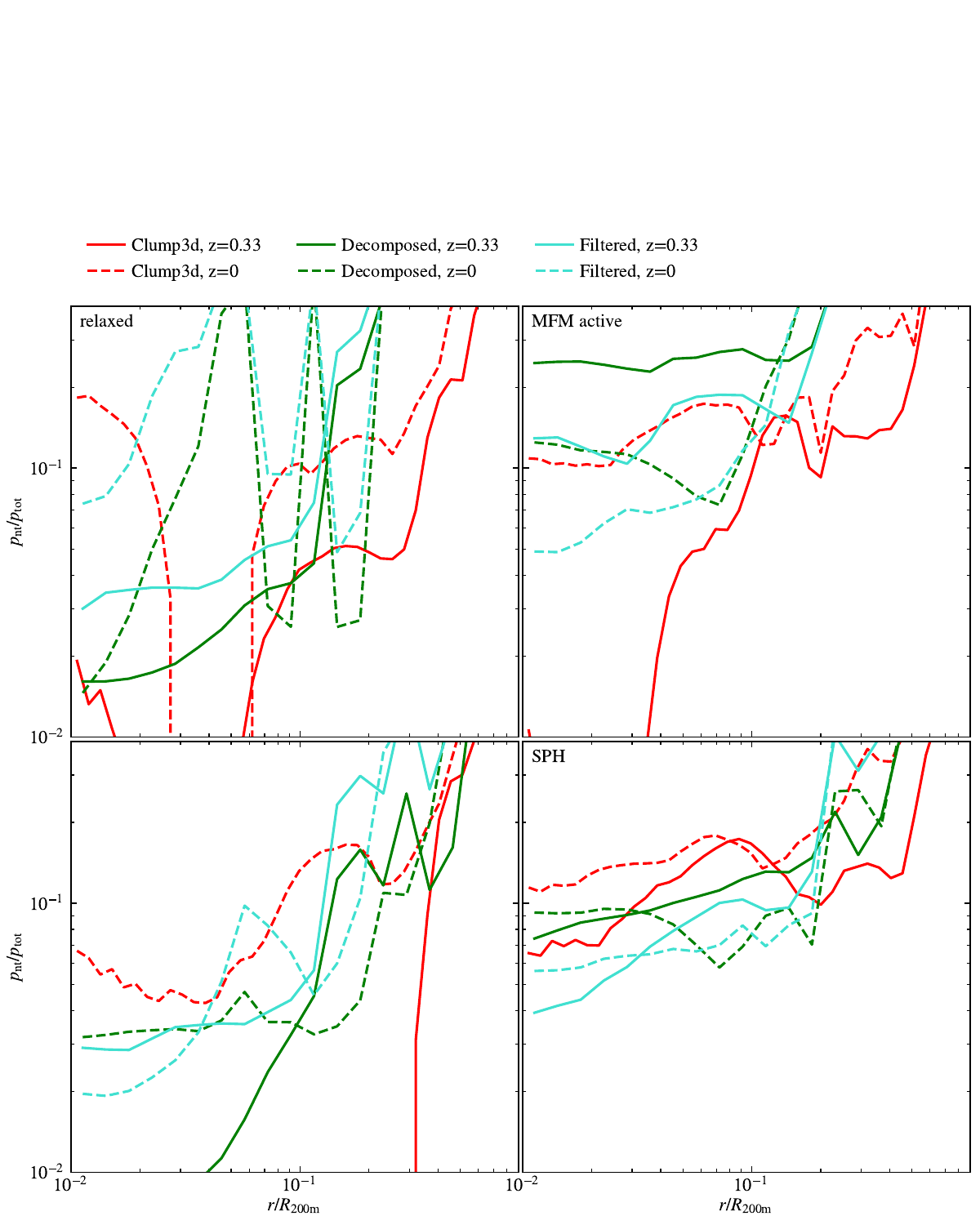}
    \caption{Turbulent pressure profile averaged over all clusters at redshift $z=0.33$ and $z=0$, comparing the three analysis methods: the clump3d method, the solenoidal velocity component, and the multi-scale filtered velocity. The sample is split between dynamical states (left column: relaxed, right column: active) and hydro-methods (top row: MFM, bottom row: SPH) used for the simulation.
    The linestyle indicates the redshift, the color the analysis method.
    As only seven clusters are used for averaging, a strong scatter between individual clusters dominates the uncertainty.}
    \label{fig:Pturb_combined_redshift}
\end{figure*}

Finally, we can use the methods described in Sec.~\ref{sec:analysis_methods} to derive non-thermal/turbulent pressure fractions.
We start by showing values at two individual redshifts, i.e. at $z=0$ and $z=0.33$, in Fig.~\ref{fig:Pturb_combined_redshift}, to better compare to observations that typically cover a narrow redshift range.

A strong scatter is present as the number of clusters used for averaging is small. Only two of the clusters are relaxed at redshift $z=0$, and five of them are active.
Single outliers and timings of ongoing mergers can significantly influence the value. We find an increase in non-thermal pressure towards the outskirts, where the cluster is not in equilibrium but dominated by bulk, turbulent motions, and the presence of a larger number of substructures.

All the methods predict a non-zero turbulent pressure fraction with only a few exceptions. Due to the strong scatter, no clear differences are visible between the individual analysis methods.
In addition, there are no clear trends for the turbulent pressure fraction with redshift.
We find some increase in turbulent pressure for active clusters compared to relaxed ones, where this seems to be even stronger for MFM compared to SPH. Relaxed clusters have on average only a few percent of turbulent pressure support, while active ones can have up to $\approx 20\%$.
Nevertheless, these trends are not significant and all non-thermal pressure profiles are consistent within the strong scatter among individual clusters. 

\begin{figure*}
    \centering
    \includegraphics[width=\linewidth]{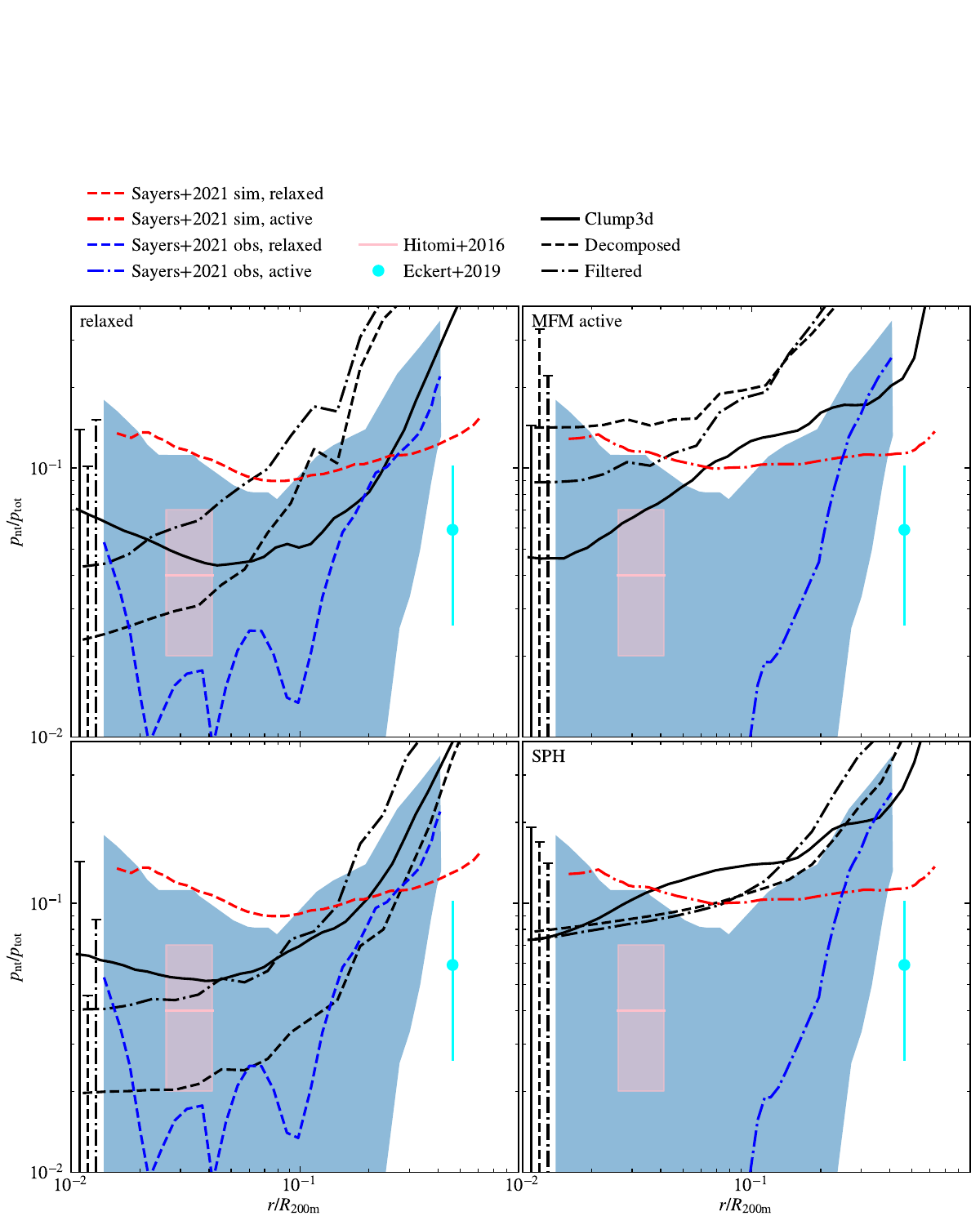}
    \caption{Turbulent pressure profile averaged over all clusters and redshifts $0.43\geq z\geq0$. The sample is split between dynamical states and hydro-methods used for the simulation, shown in the different panels. The solid line shows the results for the Clump3d analysis, the dashed line the pressure resulting from the solenoidal velocity component using the Helmholtz-decomposed velocity, and the dash-dotted line results from the multi-scale filtering.\\ The typical uncertainty is on the order of $\sigma=0.08$ and indicated for each method with an error bar on the left of each panel.}
    \label{fig:Pturb_combined}
\end{figure*}

A clearer picture arises when averaging over redshifts $0.43\geq z\geq0$, as shown in Fig.~\ref{fig:Pturb_combined}.
We include different data from the literature as a comparison, including X-ray observations in the Perseus cluster \citep{HitomiCollaboration+2016,HitomiCollaboration+2018}, averages from the X-COP sample \citep{Eckert+2019}, as well as the observations and ``The 300 simulations'' \citep{Cui+2018} analyzed by \citet{Sayers+2021}. These span a wide range of possible turbulent pressure fractions from $P_{\rm nt}/P_{\rm tot}=0$ up to $\approx0.13$.

All the methods predict an increase in the outskirts located around $20-40\%~R_{200\text{m}}$, consistent with the observations by \citet{Sayers+2021}, but stronger than for their simulation results. 
We typically find pressure values greater than zero, but still consistent with zero within the cluster-to-cluster and redshift-to-redshift scatter, which is typically around $0.08$. The average $1\sigma$ scatter within $0.1 \, R_{200}$ for each method is indicated by a bar on the left of each panel. 

For the relaxed clusters, we find turbulent pressure values smaller than ``The 300 simulations'' analyzed by \citet{Sayers+2021} within the central region. 
Differences can be attributed to not including feedback processes in our simulations, which could potentially increase the amount of turbulence. Our results should thus be considered as lower limits.
We find that the relaxed sub-sample is consistent with the spectral observations in the Perseus cluster \citep{HitomiCollaboration+2016,HitomiCollaboration+2018}, which is often assumed to be relaxed in the center, even though showing tracers of activity in the outskirts.

Active clusters yield higher turbulent pressure values, more consistent with the simulations of the active clusters by \citet{Sayers+2021}.
In general, our predictions, even though not including feedback processes, lie within the expected range of values found by other simulations and observations.

The clump3d method analyzing the deviation from HE predicts a non-thermal pressure around $\lesssim10\%$. It does not lie above the other two methods: rather, it lies below their predictions for MFM. This is opposed to the aim of the technique, which should indicate an upper limit for turbulence, as it also includes the non-thermal pressure contribution from bulk motions.
A mild increase for active clusters compared to relaxed ones is present both for MFM and even stronger for SPH.
Overall, the differences between hydro-methods are very small.

In contrast, the velocity-based methods show much clearer trends with the hydro-method and dynamical state. Interestingly, despite the radial velocity profiles showing a higher solenoidal velocity than the filtered one, we find the opposite trend for the pressure evaluated from the squared velocity. Nevertheless, the two methods produce very similar results. 
Relaxed clusters have a turbulent pressure fraction in the center of $2-4\%$. We find a strong increase in turbulent pressure for active clusters to $8-9\%$ for SPH, and even higher up to $9-13\%$ for the MFM simulation.

Overall, we find small, but non-negligible turbulent pressure fractions in all simulations, which are consistent with previous results.

\subsection{Velocity distributions via line profiles} \label{sec:line_profiles}

An alternative method to analyze the distribution of velocities is via line profiles, which also offers a more direct comparison to observational results.
We follow a simplified approach, as described by \citet{Dolag+2005b}. When evaluating the line profile, the thermal broadening is ignored, and we focus purely on Doppler broadening due to velocity.

We assume a constant iron abundance and emissivity proportional to the density, independent of temperature
\begin{align}
    \epsilon\sim~& n_e^2\Delta V\sim \rho.
\end{align}
The total emission is computed as the sum of all particles $i$ within the virial radius and a cylinder in the direction of the sightline in $z$-direction with radius $r=150~h^{-1}$kpc
\begin{align}
    I\left(\Delta E\right) =~& \sum_{i\in V} \epsilon_i\delta\left(\Delta E_i\right), \\
    \Delta E_i =~& E_0\frac{\sqrt{1-\beta_i^2}}{\left(1+\beta_i\cos\phi_i\right)}.
\end{align}
The energy of the line is shifted due to the relativistic Doppler shift with $\beta_i=\abs{\myvector{v}_i}/c$ and angle of the velocity with respect to the line of sight $\cos\phi_i=\abs{\myvector{v}_{{\rm los},i}}/\abs{\myvector{v}_i}$. As we neglect any thermal broadening, each individual line is described by a Dirac delta function $\delta$.
The impact parameter is varied between $0$, $250~h^{-1}$kpc and $500~h^{-1}$kpc along the $x$-direction.

In Fig.~\ref{fig:line_profiles}, we show the corresponding profiles binned at a resolution of $1$~eV for a $6.702$~keV iron line for all our clusters. 
As the density is highest in the center, the intensity of the line profile looking through the cluster center is the highest, and it decreases as the profile is evaluated moving further from the galaxy cluster center.

All line profiles are significantly broadened due to bulk- and turbulent motions by several $10$~eV. Their shape can vary between close to Gaussian to highly irregular profiles, depending on the velocity structure.
Clusters g14, g19, and g55, which have the highest turbulent velocities in the radial profiles shown in Fig.~\ref{fig:velocity}, also show the broadest and most irregular line profiles compared to more relaxed clusters broadened by more than $50$\,eV.

Comparing the ratio of emissivity with MFM to SPH, we find that for many clusters, this ratio increases towards the wings of the line, which indicates that the distribution is broader for MFM compared to SPH.
For several cases, this can be seen even from the line profile itself.
An exception is cluster g63, where a secondary maximum appears for SPH, but not MFM. This is most likely connected to a substructure at that position. Also \citet{Dolag+2005b} argue that this method is highly sensitive to the timing and position of substructures.

Overall, the line profiles confirm the previous findings that more turbulence is present for MFM compared to SPH. Even if the line profile includes information not only on the turbulent motion but also on bulk velocities, it is a good tracer of turbulence.

\section{Conclusions} \label{sec:conclusions}

We have analyzed the turbulent pressure support in galaxy clusters simulated with different hydro-methods, where the turbulence has been estimated in three different ways.
Our set of zoom-in regions includes clusters at various dynamical states, so that even using few clusters we can get meaningful results.

The amount of turbulent pressure can vary significantly depending on the analysis method which is employed.
The multi-scale filtering is the most direct approach, as it filters out bulk motions depending on the local structure.
Like the filtered velocity, also the solenoidal velocity is a tracer of the turbulent velocity. 
Compared to the filtered velocity, it is slightly less direct and shows some differences in radial velocity profiles and derived turbulent pressure. Also other solenoidal motions can be present, and turbulence can in principle also have compressive components, even though small in the ICM. Despite these limitations, we find the solenoidal component to follow the results based on the multi-scale filtered velocity reasonably well, especially considering that it is much computationally less expensive to calculate.
While the clump3d method should predict an upper limit for the turbulence as it includes the effect of bulk motions, we find values similar to or even lower than those of the velocity-based methods.
Possible explanations are limitations of the clump3d method, which relies on several assumptions, or that not all solenoidal or small-scale motion acts as an actual source of pressure for the HE equation.
To unquestionably assess the reason for this discrepancy, a more detailed analysis beyond the scope of this paper is necessary. 

Our setup allows us to compare hydro-methods for simulating turbulence in the ICM beyond idealized simulations, for the first time.
The global structure is very similar between MFM and SPH, with differences on smaller scales. 
Visual inspection of the projected surface density reveals more small-scale density fluctuations for MFM compared to SPH. This can partly be explained by the effectively higher resolution for MFM, but also the better capturing of subsonic turbulence via the MFM scheme. Even though the reduced amount of small substructures for MFM is not statistically significant, this indicates more numerical dissipation and consequently more mixing for MFM.

Higher turbulent velocities are present for MFM than for SPH.
Consistent with more idealized simulations, MFM predicts more turbulence for the velocity-based methods. For the clump3d method, only minor differences are present.

Finally, we analyzed the impact of the dynamical state.
By exploiting velocity-based methods, we find that active clusters have more turbulence than relaxed ones, in contrast to \citet{Sayers+2021}.
Some differences with respect to their work can be attributed to feedback processes, which are not present in our simulations as we instead preferred to have a clearer setup. Stellar and AGN feedback can drive feedback on small scales, increasing the overall amount of turbulence. While AGN feedback affects the cluster mainly in the center, the effect of stellar feedback can be more widespread.
Although not including feedback processes, the amount of turbulence we find in our simulations is consistent with the range of values found in previous works between a few percent for relaxed clusters and up to $\approx 13\%$ in the center for active clusters.

As the turbulent pressure increases towards the outskirts, the hydrostatic bias at the X-ray boundary around $R_{500}\approx0.7R_{200}$, where the signal is obtained by observations, is small, but non-negligible.
We don't expect this result to change due to feedback processes.

As a final remark, we stress that it is key to quantify which are the discrepancies among different simulations and analysis techniques, and why they arise: this allows us to understand their (possibly different) predicted non-thermal or turbulent pressure support.

\subsection{Outlook}
While this work focused on a clean setup using purely hydrodynamical simulations, the inclusion of additional physical processes might partly change the picture. Especially, cooling would act as a sink for energy. Star formation, stellar, and AGN feedback in contrast would act as small-scale drivers of turbulence. 
As feedback processes self-regulate, they might also reduce differences between hydro-methods. More studies including feedback processes would be necessary to confirm our findings beyond non-radiative simulations.
In addition, this could give further insight into the expected amount of turbulence in real galaxy clusters.

Also the inclusion of magnetic fields would provide an additional channel to study turbulence, as they closely couple via the dynamo effect. In addition, they can change the shape of turbulence and leave an imprint on the power spectrum.

\begin{acknowledgements}
We thank the anonymous referee for the constructive feedback which helped to improve the quality of this paper.
FG and KD acknowledge support by the COMPLEX project from the European Research Council (ERC) under the European Union’s Horizon 2020 research and innovation program grant agreement ERC-2019-AdG 882679. 
MV acknowledges support by the Italian Research Center on High Performance Computing, Big Data and Quantum Computing (ICSC), project funded by European Union - NextGenerationEU - and National Recovery and Resilience Plan (NRRP) - Mission 4 Component 2, within the activities of Spoke 3, Astrophysics and Cosmos Observations, and by the INFN Indark Grant.
UPS is supported by a Flatiron Research Fellowship at the Center for Computational Astrophysics (CCA) of the Flatiron Institute. The Flatiron Institute is supported by the Simons Foundation. FG, MV, and KD acknowledge support by the Deutsche Forschungsgemeinschaft (DFG, German Research Foundation) under Germany’s Excellence Strategy - EXC-2094 - 390783311. MV is supported by the Alexander von Humboldt Stiftung and the Carl Friedrich von Siemens Stiftung. We are especially grateful for the support by M. Petkova through the Computational Center for Particle and Astrophysics (C2PAP) under the project pn68va. Some simulations were carried out at the Leibniz Supercomputer Center (LRZ) under the project pr86re (SuperCast). The analysis was performed mainly in julia \citep{Bezanson+2014}, including the package GadgetIO by \citet{GadgetIO}.
DVP has been supported by the Agencia Estatal de Investigación Española (AEI; grant PID2022-138855NB-C33), by the Ministerio de Ciencia e Innovación (MCIN) within the Plan de Recuperación, Transformación y Resiliencia del Gobierno de España through the project ASFAE/2022/001, with funding from European Union NextGenerationEU (PRTR-C17.I1), by the Generalitat Valenciana (grant CIPROM/2022/49), and by Universitat de València through an Atracció de Talent fellowship.
\end{acknowledgements}

\bibliographystyle{aa} 
\bibliography{quellen}

\begin{appendix}

\section{Line profiles}
\begin{figure*}
    \centering
    \includegraphics[width=\textwidth]{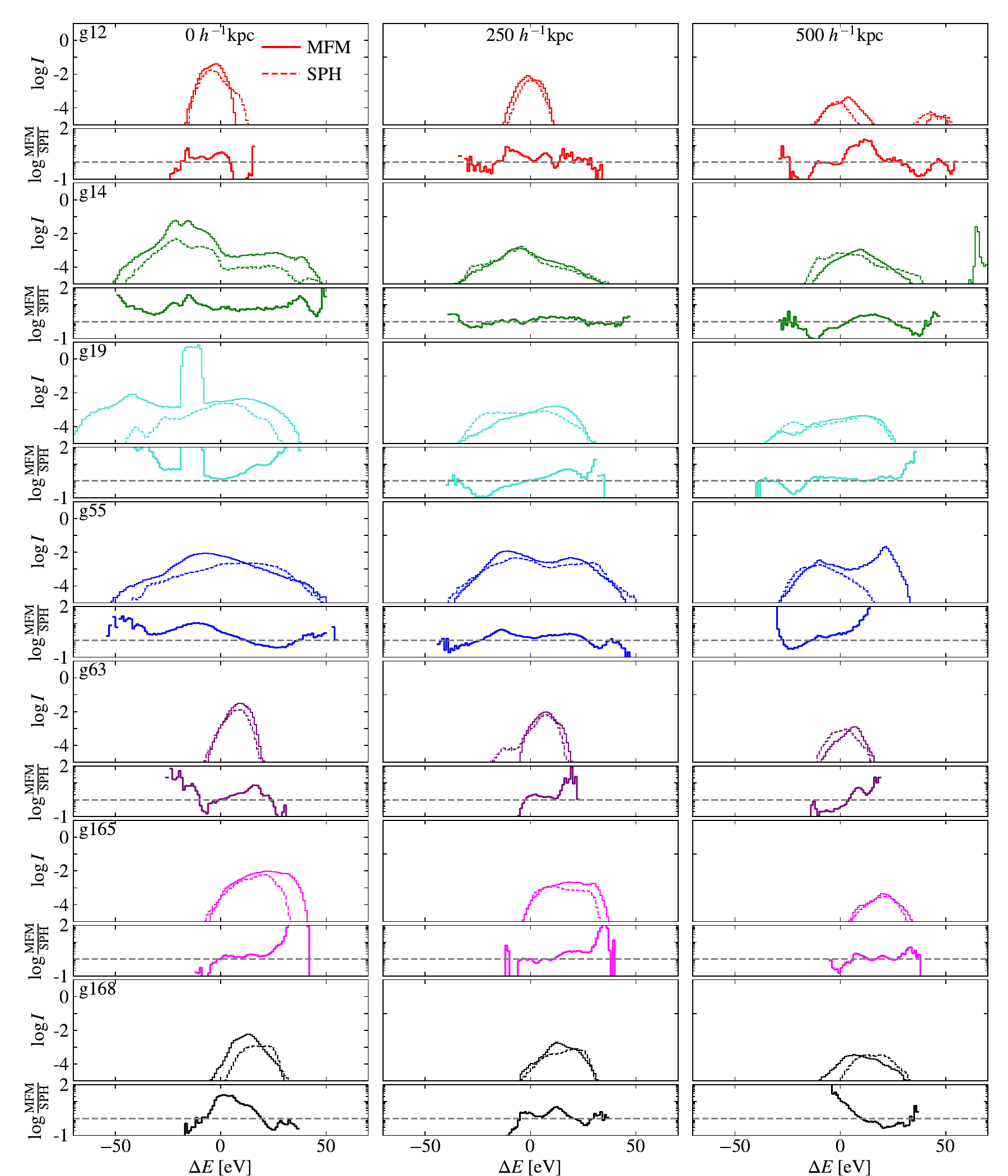}
    \caption{Line profiles of a $6.702$~keV iron line for the different clusters at different distances from the cluster center (as written in each of the top panels). The intensity is in arbitrary units. For each panel, we also calculate the ratio between intensity for MFM and SPH to emphasize smaller differences.}
    \label{fig:line_profiles}
\end{figure*}

In Fig.~\ref{fig:line_profiles} we show the line profiles at three different distances from the cluster center of all clusters. See Sec.~\ref{sec:line_profiles} for more details.

\section{Surface density} \label{app:surface_density_plots}

\begin{figure*}
    \centering
    \includegraphics[width=0.995\textwidth]{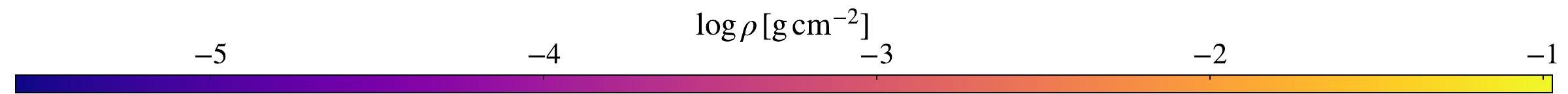}
    \includegraphics[width=\columnwidth]{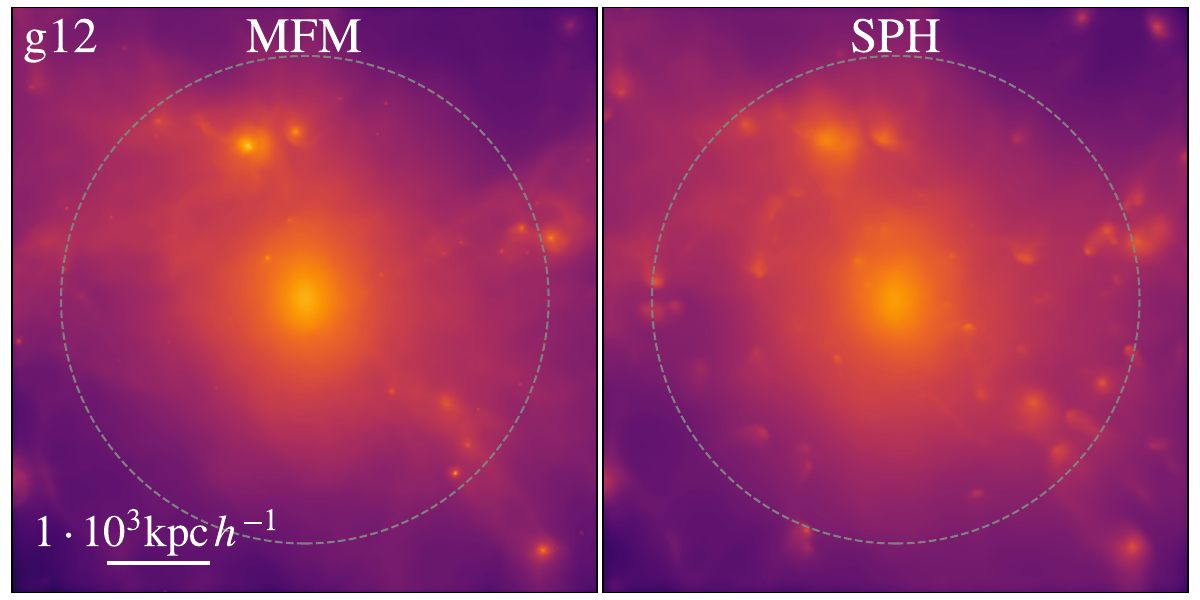}
    \includegraphics[width=\columnwidth]{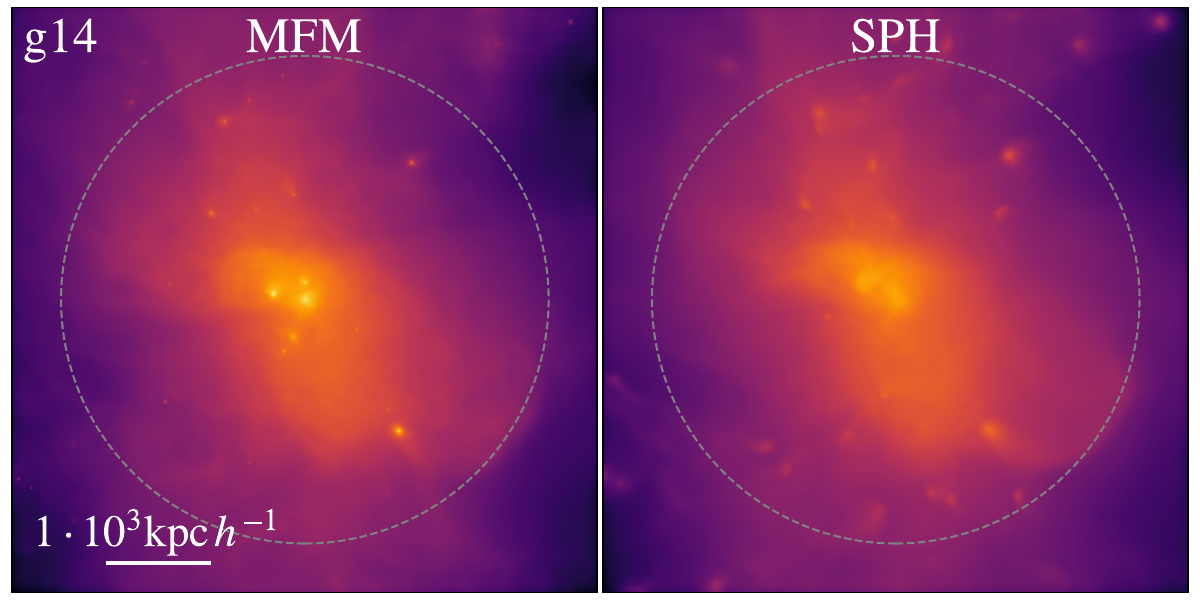}
    \includegraphics[width=\columnwidth]{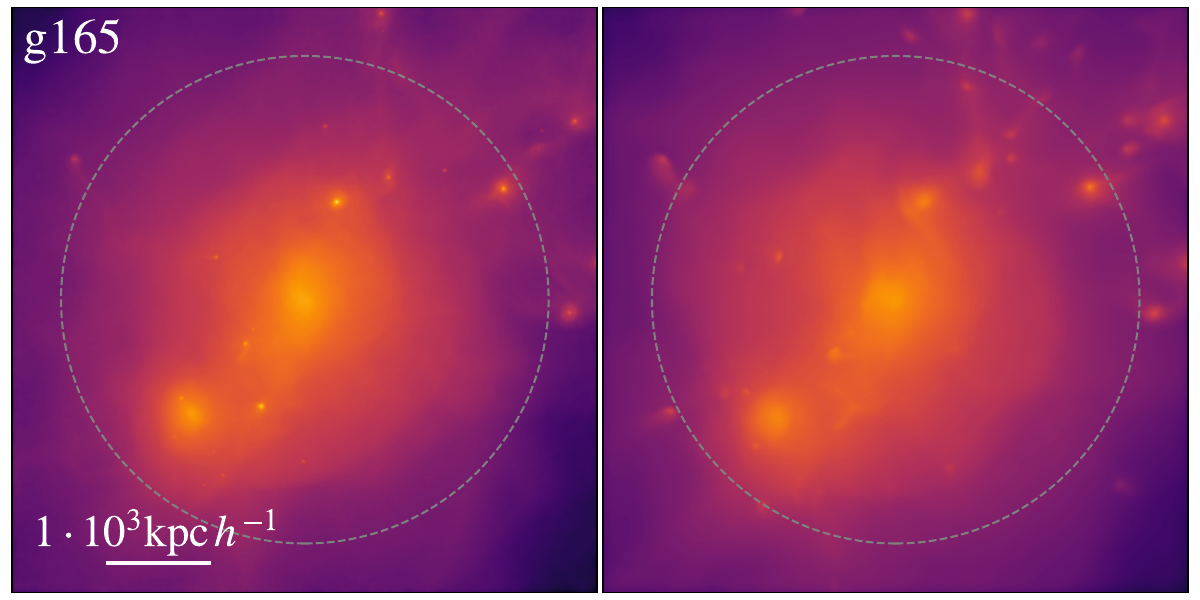}
    \includegraphics[width=\columnwidth]{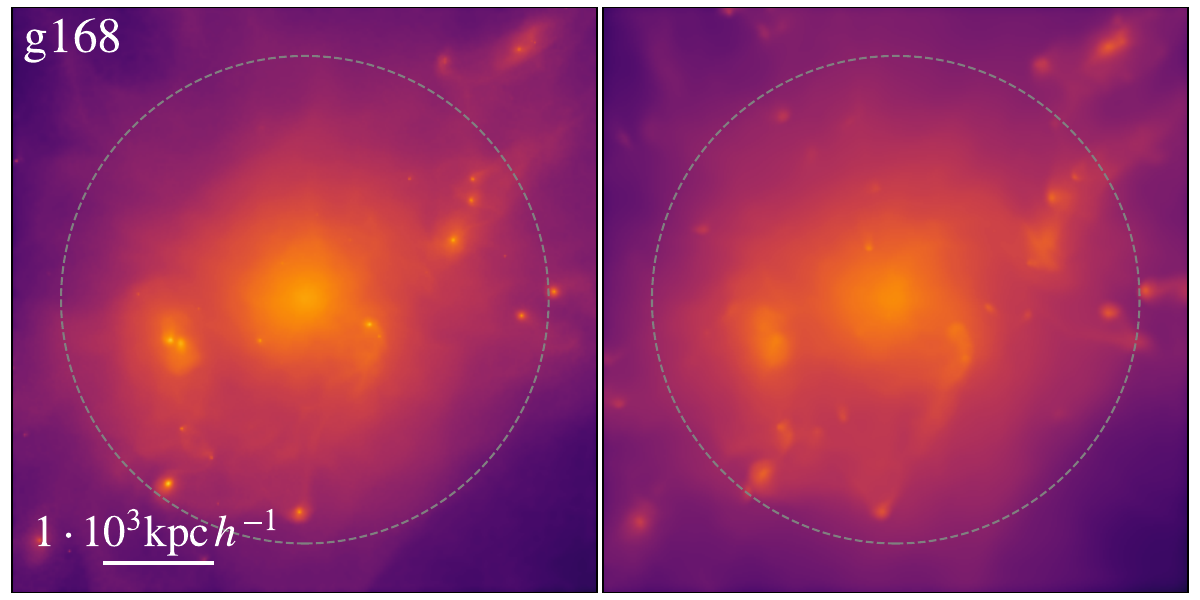}
    \includegraphics[width=\columnwidth]{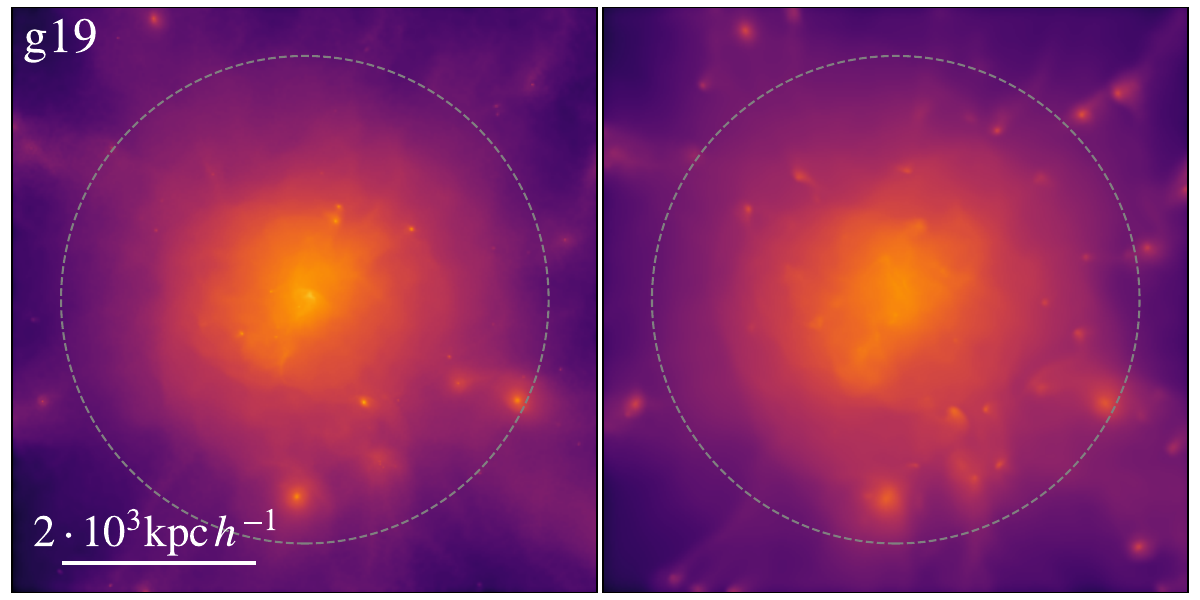}
    \includegraphics[width=\columnwidth]{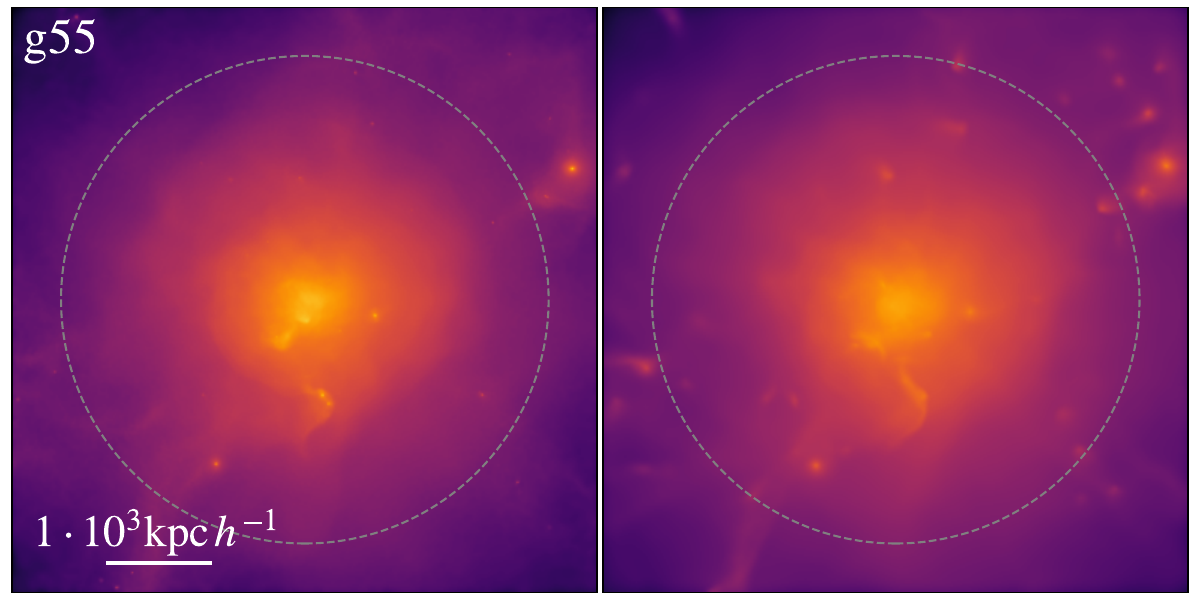}
    \includegraphics[width=\columnwidth]{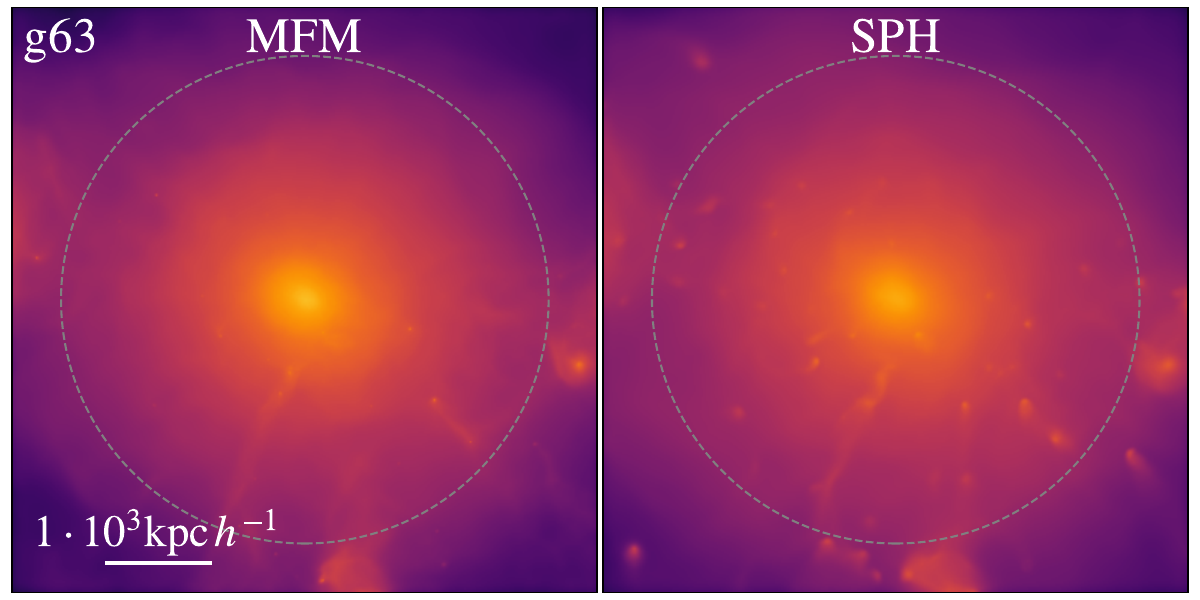}
    \caption{Projected gas density maps for all clusters analyzed in this work at redshift $z=0$. The dashed circle denotes $R_{\rm vir}$.}
    \label{fig:all_maps}
\end{figure*}

In Fig.~\ref{fig:all_maps} we show the surface density of all clusters.
More active clusters such as g14 or g19 show a lot of substructure, while more relaxed clusters such as g63 show much less substructure. In g14 even two distinct cores in the center are visible as tracers of a recent major merger.

\end{appendix}
\end{document}